\documentclass[
prd,
superscriptaddress,
10 pt,
twocolumn,
preprintnumbers,
notitlepage,
secnumarabic,
nobibnotes,
nofootinbib,
showpacs
]{revtex4-1}

\usepackage[T1]{fontenc}
\usepackage[pdftex]{color,graphicx}
\usepackage{mathtools}
\usepackage{amssymb}
\usepackage{bm}
\usepackage{dsfont}
\usepackage{mathrsfs}
\usepackage{calligra}
\usepackage{tabularx}
\usepackage[artemisia]{textgreek}
\usepackage{hyperref}
\usepackage{xcolor}
\usepackage{enumerate}
\usepackage{multirow}
\usepackage{url}
\usepackage{caption}
\usepackage{ulem}

\usepackage{perpage}
\MakePerPage{footnote}
\renewcommand{\.}{\!\;}		
\renewcommand{\@}{\!\:\!}	

\DeclareMathAlphabet\mathbfcal{OMS}{cmsy}{b}{n}

\renewcommand{\i}{\!\:\mathrm{i}}

\newcommand{\pos}[1]{\ensuremath{\@\langle#1\rangle}}

\begin{document}
\title{Implementation of the holonomy representation
\\
of the Ashtekar connection in loop quantum gravity}

\author{Jakub Bilski}
\email{bilski@zjut.edu.cn}
\affiliation{Institute for Theoretical Physics and Cosmology, Zhejiang University of Technology, 310023 Hangzhou, China}


\begin{abstract}
\noindent
The improved lattice regularization method of the Ashtekar connection holonomy representation in loop quantum gravity is described in this article. The approach is based on the geometric expansion of holonomies into power series up to the quadratic order terms in the regularization parameter. As a result, a more accurate procedure than the currently established approach to the canonical lattice quantization of gravity is obtained. Moreover, if holonomies are defined along linear links, this procedure becomes exact. Furthermore, in the improved method, the symmetry of holonomies assigned to links is directly reflected in the related distribution of connections. Finally, the domain of the lattice-regularized Hamiltonian constraint takes a natural structure of elementary cells sum. Consequently, under certain restrictions, the related scalar constraint operator, which spectrum is independent of intertwiners, can be defined.
\end{abstract}

\maketitle


\section{Introduction}\label{Sec_Introduction}

\noindent
The notion of loop quantum gravity (LQG) refers to a collection of models \cite{Ashtekar:2017yom} originally formulated to quantize the gravitational field in a rigorous accordance with the general postulate of relativity \cite{Einstein:1916vd}. This is revealed in the background independence and the nonperturbative construction. However, all of these approaches assume the time gauge implementation \cite{Arnowitt:1960es}. A common denominator of these different theories is the selection of the Ashtekar variables \cite{Ashtekar:1986yd} and the lattice regularization method in which gauge connections are replaced by holonomies.

The first complete construction of the theory, called canonical LQG, was introduced in \cite{Thiemann:1996ay,Thiemann:1996aw} and it is the most developed formulation as of today. Its modern variant \cite{Thiemann:2007zz} assumes the regularization method of the real Ashtekar-Barbero connection and its curvature \cite{Barbero:1994ap} that expresses these objects as particular functionals of holonomies. In this article, this established procedure is going to be verified and some aspects concerning the regularization procedure are questioned. Then, their more accurate and more consistent formulation is proposed.


\section{Regularization of the Hamiltonian constraint}\label{Sec_Regularization}

\noindent
The standard analysis, leading toward the canonical quantization of the gravitational field, initiates from postulating the Holst action \cite{Holst:1995pc}. The Legendre transform leads to three types of first-class constraints. Two of them, the one describing spatial diffeomorphisms and the one implementing the $\mathfrak{su}(2)$ invariance, are solved classically. The third one, called the Hamiltonian constraint,
\begin{align}
\label{A_scalar}
\!\!\int\!\!\frac{d^3x}{\kappa}N
|E|^{-\frac{1}{2}}\big(F^i_{ab}-(\gamma^2\@+1)\epsilon_{ilm}K^l_aK^m_b\big)\epsilon^{ijk}E_j^aE_k^b,\!\!
\end{align}
carries the propagating gravitational degrees of freedom. Here, $\kappa:=16\pi G/c^4$ is the Einstein constant. The lapse function $N$ is a Legendre multiplier and $\gamma$ denotes the Immirzi parameter. The dynamical quantities are: the densitized dreibein $E^a_i$, where $E$ represents its determinant, the dreibein-contracted extrinsic curvature $K^i_a$, and the curvature of the Ashtekar connection $F^i_{ab}:=\partial_aA^i_b-\partial_bA^i_b+\epsilon^{ijk}A^j_aA^k_b$. The two sets of indices, $a,b,...$ and $i,j,...$, denote the external spatial and the internal gauge directions in three-dimensional spaces, respectively.

Before quantization, the expression in \eqref{A_scalar}, being a sum of two significantly different elements, has to be properly regularized. The second term, multiplied by the factor $\gamma^2+1$, can be expressed as a functional of the first one --- see \cite{Thiemann:1996ay,Thiemann:1996aw,Thiemann:2007zz}. To simplify this article, the recollection of this long result is not necessary, but it is worth to emphasize that its construction does not require any approximation.

The first term, proportional to the quantity $|E|^{-\frac{1}{2}}F^i_{ab}\epsilon^{ijk}E_j^aE_k^b$, is regularized with an approximate method. This method consists of two steps. The first one is exact. It is based on the identity introduced in \cite{Thiemann:1996ay}, given by the formula
\begin{align}
\label{trick_gravity_TT}
\sigma|E|^{-\frac{1}{2}}\epsilon_{ijk}E^b_jE^c_k
=\frac{4}{\gamma\kappa}\epsilon^{abc}\big\{\mathbf{V}(R),A^i_a\big\}\,,
\end{align}
where the Poisson brackets are regarding the canonical ADM variables $q_{ab}$ and $p^{ab} $ (in the given order) \cite{Arnowitt:1960es}. The new quantities are the parameter $\sigma:=\text{sgn}(\text{det}(E^a_i))$ and the volume of a region $R$
\begin{align}
\label{volume}
\mathbf{V}(R):=\!\int_{\!R}\!\!d^3x\sqrt{|E(x)|}\,.
\end{align}

The second step requires the introduction of the holonomy of the Ashtekar connection $A_a:=A_a^i\tau_i$ ($\tau_i$ is the $\mathfrak{su}(2)$ generator, satisfying $[\tau_j,\tau_k]=\epsilon_{ijk}\tau_i$), given by the expression
\begin{align}
\label{holonomy}
h_{p}^{-1}\pos{v}:=\mathcal{P}\exp\!\bigg(\!-\!\int_{\@l_p\pos{v}}\!\!\!\!\!\!\!ds\,\dot{\ell}^a\@(s)\.A_a\@\big(\ell(s)\big)\!\bigg).
\end{align}
In this formula, the holonomy is defined along a smooth path ${l_{p\.}\pos{v}}:={\mathbb{L}_0\varepsilon_{p\.}\pos{v}}$ identified with a graph's edge between vertices $v$ and $v+l_p$. Here, $\mathbb{L}_0$ is a fiducial length scale, while ${\varepsilon_{p\.}\pos{v}}$ denotes the dimensionless regularization parameter corresponding to a particular edge. Moreover, the reciprocal LQG's convention \cite{Ashtekar:2017yom,Thiemann:2007zz}, in which one denotes a holonomy by $h^{-1}$, has been applied. In what follows, the holonomies-dressed graph is going to be called a lattice, its edges, links, and the points of their intersections, nodes. In this article the lattice, in which the closed paths composed of a few links are triangles, is considered. This structure reflects the so-called triangulation\footnote{More precisely, assuming that the links of the lattice are not linear but analytical paths, the partition of the manifold should be called a pseudotriangulation \cite{Taylor:2004}. In this case, the faces of elementary tetrahedral cells are pseudotriangles. They are the models of triangular faces deformed in non-Euclidean geometry.} of the Cauchy hypersurface, i.e. the submanifold indicated by the time gauge introduction \cite{Arnowitt:1960es} on the manifold in which the Holst action is defined (see \cite{Thiemann:1996aw,Thiemann:2007zz}). This form of the lattice is usually studied in the context of LQG. However, any non-Euclidean manifold with a defined geometry can be tessellated, i.e. divided into solids with polygon faces. Thus, the lattice with Euclidean triangular, quadrilateral, or more complicated loops can be considered as well \cite{Bilski:2021hrr}.


\section{Standard implementation}\label{Sec_Standard}

\noindent
The lattice-regularization of the Ashtekar connection is an approximate method based on the holonomy power series expansion
\begin{align}
\begin{split}
\label{holonomy_expansion}
\!h_{(\@p\@)}^{\mp1}\pos{v}=&\;\mathds{1}\mp\mathbb{L}_0\.\varepsilon_{(\@p\@)}\pos{v}A_{(\@p\@)}\pos{v}
\mp\frac{1}{2}\mathbb{L}_0^2\.\varepsilon_{(\@p\@)}^2\pos{v}\.\partial_{(\@p\@)}A_{(\@p\@)}\pos{v}\!\!\!
\\
&+\frac{1}{2}\mathbb{L}_0^2\.\varepsilon_{(\@p\@)}^2\pos{v}A_{(\@p\@)}\pos{v}A_{(\@p\@)}\pos{v}
+\mathcal{O}(\varepsilon^3)\,.
\end{split}
\end{align}
This expansion was performed around the infinitesimal value of the dimensionless regulator $\varepsilon_p:=\varepsilon_{l_p}$. In this article, the convention, in which the indices written in brackets are not summed, is applied.

Let $\alpha$ denote a direction from the endpoint of an oriented link $l_q$ toward the endpoint of $l_r$. By expansion of the holonomy around a triangular loop,
\begin{align}
\begin{split}
\label{loop_holonomy_linear}
h_{(\@q\@)\@(\@r\@)}\pos{v}:=&\;
h^{\mathstrut}_{(\@q\@)}\pos{v}\.
h^{\mathstrut}_{(\@\alpha\@)}\pos{v\@+\@l_{(\@q\@)}}\.
h^{-1\mathstrut}_{(\@r\@)}\pos{v}
\\
=&\;
h^{\mathstrut}_{(\@q\@)}\pos{v}\.
h^{-1\mathstrut}_{(\@\alpha\@)}\pos{v\@+\@l_{(\@r\@)}}\.
h^{-1\mathstrut}_{(\@r\@)}\pos{v}\,,
\end{split}
\end{align}
one can formulate a significantly different relation than the one in \eqref{holonomy_expansion},
\begin{align}
\label{loop_holonomy_expansion}
h_{(\@q\@)\@(\@r\@)}\pos{v}=\mathds{1}
+\frac{1}{2}\mathbb{L}_0^2\.\varepsilon_{(\@q\@)}\pos{v}\.\varepsilon_{(\@r\@)}\pos{v}F_{(\@q\@)\@(\@r\@)}\pos{v}+\mathcal{O}(\varepsilon^3)\,.
\end{align}
One could presume that perhaps this difference was the reason behind formulating the original regularization of LQG \cite{Thiemann:1996ay,Thiemann:1996aw} in the way involving the terms up to the order $\varepsilon^2$ in expression \eqref{loop_holonomy_expansion} but only up to the order $\varepsilon$ in formula \eqref{holonomy_expansion}.

By following this original idea of an unbalanced approximation, let three order-$\varepsilon$ relations be proposed. These subsequent expressions are constructed by inverting the holonomy expansion in \eqref{holonomy_expansion} into the functionals of a connection given in terms of holonomies,
\begin{align}
\label{Poisson_holonomy_1}
\begin{split}
\big\{h_{(\@p\@)}\pos{v},\text{f}[E^a\pos{v}]\big\}
=&\;l_{(\@p\@)}\pos{v}\big\{A_{(\@p\@)}\pos{v},\text{f}[E^a\pos{v}]\big\}
\\
&+\mathcal{O}(\varepsilon^2)\,,
\end{split}
\\
\label{Poisson_holonomy_2}
\begin{split}
\!\!h_{(\@p\@)}^{-1\mathstrut}\pos{v}\big\{h^{\mathstrut}_{(\@p\@)}\pos{v},\text{f}[E^a\pos{v}]\big\}
=&\;l_{(\@p\@)}\pos{v}\big\{A_{(\@p\@)}\pos{v},\text{f}[E^a\pos{v}]\big\}\!\!
\\
&+\mathcal{O}\big(\varepsilon^2\big)\,,
\end{split}
\\
\label{Poisson_holonomy_3}
\begin{split}
\!\!\!2^{-n+1}\big\{h_{(\@p\@)}^n\pos{v},\text{f}[E^a\pos{v}]\big\}
=&\;l_{(\@p\@)}\pos{v}\big\{A_{(\@p\@)}\pos{v},\text{f}[E^a\pos{v}]\big\}\!\!\!
\\
&+\mathcal{O}(\varepsilon^2)\,,
\qquad
n\in\mathds{Z}_+\,.
\end{split}
\end{align}
It should be pointed out that the expression in \eqref{Poisson_holonomy_3} is a generalization of the simplest formula in \eqref{Poisson_holonomy_1} and $\text{f}[E^a\pos{v}]$ denotes any functional of the densitized dreibein localized at (or smeared around) $v$. It is also worth mentioning that the relation in \eqref{Poisson_holonomy_2} is a standard expression, which is typically selected to smear the Ashtekar connection degrees of freedom over the lattice in the currently established formulation of LQG \cite{Thiemann:1996ay,Thiemann:1996aw,Thiemann:2007zz}.

Then, in the regularization procedure of the Hamiltonian constraint, the Poisson brackets on the right-hand side of equation \eqref{trick_gravity_TT} become replaced by the Poisson brackets of a volume and a functional of holonomies. The remaining variables in \eqref{A_scalar} are assigned to the lattice by the means of using the symmetrized version of the relation in \eqref{loop_holonomy_expansion},
\begin{align}
\label{loop_holonomy_symmetrized}
\begin{split}
h_{(\@q\@)\@(\@r\@)}\pos{v}-h_{(\@q\@)\@(\@r\@)}^{-1}\pos{v}
=&\;\mathbb{L}_0^2\.\varepsilon_{(\@q\@)}\pos{v}\.\varepsilon_{(\@r\@)}\pos{v}F_{(\@q\@)\@(\@r\@)}\pos{v}
\\
&+\mathcal{O}(\varepsilon^3)\,.
\end{split}
\end{align}

It should be indicated that any of the formulas in \eqref{Poisson_holonomy_2}, \eqref{Poisson_holonomy_3}, or their appropriately constructed combinations, can be used (together with the relation in \eqref{loop_holonomy_symmetrized}) to provide a lattice-regularized representation of the Hamiltonian constraint in \eqref{A_scalar}. Precisely, this last quantity is regained in the limit $\varepsilon\to0$. However, by canonically quantizing these discussed formulas in the DeWitt representation \cite{DeWitt:1967yk},
\begin{align}
\label{quantization}
\begin{split}
A^i_a\to&\;\hat{A}^i_a|\ \rangle:=A^i_a|\ \rangle\,,
\\
E_i^a\to&\;\hat{E}_i^a|\ \rangle:=-\i\hbar\frac{\delta}{\delta A^i_a}|\ \rangle\,,
\\
\{X,Y\}\to&\;(\i\hbar)^{-1}[\hat{X},\hat{Y}]\,,
\end{split}
\end{align}
each result leads to a significantly different quantum theory. Let it be emphasized that these differences occur only in the structures of the Hamiltonian constraints operators. The Fock-like space \cite{Thiemann:1996aw,Thiemann:1996av,Thiemann:1997rv}, the related scalar product \cite{Ashtekar:1994mh,Ashtekar:1994wa,Ashtekar:1995zh}, and the representation of canonical operators \cite{Fleischhack:2004jc,Lewandowski:2005jk} are no different from the standard formulation of LQG \cite{Thiemann:2007zz}.

This freedom in the selection of the regularization procedure may suggest that the resulting different variants of LQG have no predictive power. Therefore, to restore the uniqueness of the quantum theory and the form of the related semiclassical corrections, it may be necessary to question all order-$\varepsilon$ holonomy-regularization relations analogous to \eqref{Poisson_holonomy_2} and \eqref{Poisson_holonomy_3}. Luckily, one can provide three arguments supporting this hypothesis.

The first one is the already mentioned inconsistency in the selection of terms in the expansions of holonomies --- up to the linear order in $\varepsilon$ in \eqref{Poisson_holonomy_2} or \eqref{Poisson_holonomy_3}, and up to the quadratic order in \eqref{loop_holonomy_symmetrized}.

The second argument is related to the transition from the continuous formulation of the Hamiltonian constraint in \eqref{A_scalar} into its lattice analog. The former quantity is defined in terms of $A^i_a(x)$, $F^i_{ab}(x)$, $K^i_a(x)$, and $E_i^a(x)$. The latter consists of ${h_{p\.}\pos{v}}:={h(l_{p\.}\pos{v})}$, ${h_{qr\.}\pos{v}}:=h\big({l_{q\.}\pos{v}}\circ{l_{\alpha\.}\pos{v+l_{(\@q\@)}}}\circ{(l_{r\.}\pos{v+l_{(\@r\@)}})^{-1}\big)}$, and $\mathbf{V}(R)$. Let $R$ be identified with a tetrahedral region bounded by four triangular faces. One can link the position of $E_i^a(x)$ with an averaging in the relation between $F^i_{ab}(x)$ and ${h_{qr\.}\pos{v}}$ by smearing the former quantity all over the volume defined in \eqref{volume}, where $x\in R$. A natural method of the mentioned averaging could be a weighted summation of the Ashtekar connection curvatures located at the four faces of $R$; these faces are indicated by the four loops around them. In this way, the point-region relation became comparable with the surface-surface boundary one. An analogous method cannot be implemented to the point-segment (or in general point-path) relation between $A^i_a(x)$ and ${h_{p\.}\pos{v}}$. In this last case, the problem concerns the fact that a holonomy defined along an edge is linked with the continuous distribution of a connection along this edge. Consequently, the relation should be formulated as a probability distribution. Therefore, its simplification can be, for instance, constructed between the holonomy along $l_{p\.}\pos{v}$ and either the connection at $v+l_p/2$ or the pair of connections at $v$ and $v+l_p$. However, the relations in \eqref{Poisson_holonomy_1},  \eqref{Poisson_holonomy_2}, and \eqref{Poisson_holonomy_3} do not reflect the symmetry of the connection's distribution along a link.

Finally, the third argument is related to the symmetry of the system after quantization. Basis states in LQG are constructed from the Wigner $D^j(h)$ matrices \cite{Wigner:1931}, which transform under the holonomy's symmetry group, namely SU$(2)$. Hence, according to Wigner's theorem \cite{Wigner:1931,Wigner:1939cj}, they must preserve the transformations of the gauge connection $A_a$ determined by the $\mathfrak{su}(2)$ representation. In other words, any SU$(2)$ holonomy must be considered at least approximately as the exponential map from the representation of some connection. The standard restriction imposed on the construction of quantum theories of gauge fields \cite{Weinberg:1995mt} requires that the expansion of the exponential map around identity is proportional to the gauge field at least up to the quadratic order in the expansion parameter. The analysis in \cite{Bilski:2020xfq} determined the conditions under which holonomies can be viewed as exponential maps from $\mathfrak{su}(2)$. Hence, the application of the expansion in \eqref{holonomy_expansion} only up to the linear terms as in \eqref{Poisson_holonomy_2} and \eqref{Poisson_holonomy_3} is not sufficiently rigorous.


\section{Improved implementation}\label{Sec_Improved}

\noindent
The accuracy of the formulation of the Ashtekar connection holonomy representation can be increased to the quadratic order in $\varepsilon$. This method will allow to properly capture the point-path relation between $A^i_a(x)$ and $h_{p\.}\pos{v}$.

One should begin investigating the right-hand side of the formula in \eqref{holonomy_expansion}. The derivative $\partial_{(\@p\@)}A_{(\@p\@)}\pos{v}$ is the limit $\varepsilon_{(\@p\@)}\pos{v}\to0$ of the difference
\begin{align}
\label{difference}
\!\frac{1}{l_{(\@p\@)}\pos{v}\!}\big(A_{(\@p\@)}\pos{v\@+\@l_{(\@p\@)}}-A_{(\@p\@)}\pos{v}\big)=\partial_{(\@p\@)}A_{(\@p\@)}\pos{v}+\mathcal{O}(\varepsilon)\,.\!
\end{align}
By applying this expression into the difference of the holonomy expansion in \eqref{holonomy_expansion} and the reciprocal expansion, this leads to
\begin{align}
\label{holonomy_difference}
\begin{split}
h_{(\@p\@)}^{\mathstrut}\pos{v}-h_{(\@p\@)}^{-1}\pos{v}=&\;l_{(\@p\@)}\pos{v}\big(A_{(\@p\@)}\pos{v}+A_{(\@p\@)}\pos{v\@+\@l_{(\@p\@)}}\big)
\\
&+\mathcal{O}(\varepsilon^3)\,.
\end{split}
\end{align}
The right-hand side of this result correctly reflects the points-path symmetry. Then, the desired refinement of the relations in \eqref{Poisson_holonomy_2} and \eqref{Poisson_holonomy_3} reads as follows:
\begin{align}
\label{holonomy_improvement}
\begin{split}
&\,\frac{1}{2}\big\{\big(h^{\mathstrut}_{p}\pos{v}-h_{p}^{-1\mathstrut}\pos{v}\big),\text{f}[E^a\pos{v}]\big\}
\\
=&\;l_{(\@p\@)}\pos{v}\Big\{
\!\!\stackrel{\scriptscriptstyle\textsc{m\!\.e\!\.a\!\.n}\!}{A}{\!\@\!}_{p}\@\big(l_{(\@p\@)}\pos{v}\big),\text{f}[E^a\pos{v}]
\Big\}
+\mathcal{O}\big(\varepsilon^3\big)\,,
\end{split}
\end{align}
where
\begin{align}
\label{mean}
\begin{split}
\!\!\stackrel{\scriptscriptstyle\textsc{m\!\.e\!\.a\!\.n}\!}{A}{\!\@\!}_{p}\@\big(l_{(\@p\@)}\pos{v}\big)
:=&\;\frac{1}{2}\big(A_{p}\pos{v}+A_{p}\pos{v\@+\@l_{(\@p\@)}}\big)
\\
\approx&\int_{\@v}^{v\@+\@l_{p}}\!\!\!\!\!\!\!\!ds\,\dot{\ell}^{q}\@(s)\.\mathcal{A}_{q}\@\big(\ell(s)\big)\,.
\end{split}
\end{align}
In the last equation, the integration has been done with the trapezoidal rule for a single interval $\mathbb{L}_0\varepsilon_{(\@p\@)}$. More precisely, the analytical path $\ell(s)$ has been approximated by a polygonal chain. Then, the integral of the Ashtekar connection's $p$-coordinate linear density $\mathcal{A}_{p}(l_{(\@p\@)})$ along the piecewise linear path has been derived in the linear approximation. This way, the continuous probability distribution of the connection has been formally calculated by assuming the trivial form of the link $l_{p\.}\pos{v}$.

As far as is known, the outcome in \eqref{holonomy_improvement} is the only geometrically-obtained expansion providing a relation between holonomies and connections, which is of order $\varepsilon^2$. Moreover, in the alternative approach, assuming from the beginning the smearing of gravitational variables over a piecewise linear lattice \cite{Bilski:2020xfq}, an even stronger relation holds. In this case, the approximation in \eqref{mean} becomes equality due to the linearity of the link $l_{p\.}\pos{v}\to\bar{l}_{p\.}\pos{v}$, where ${\forall_{\@x\in\bar{l}_{p}\pos{v}}\,A(x)=A_{(\@p\@)}\pos{v}=\,\stackrel{\scriptscriptstyle\textsc{m\!\.e\!\.a\!\.n}\!}{A}{\!\@\!}_{p}\@\big(l_{(\@p\@)}\pos{v}\big)}$.

The result in \eqref{holonomy_improvement} allows to formulate stricter and so far unique predictions from the semiclassical limit of the improving regularized lattice gravity \cite{Bilski:2021hrr}. Moreover, apart from the dedicated triple of problems in Sec.~\ref{Sec_Standard} solved at once, the structure of the left-hand side of the discussed expression has two additional advantages. It is given in the form of the Poisson brackets of the lattice-smeared objects, the positions of which are related to a single cell. Therefore, both the lattice-smeared Hamiltonian constraint's domain and the corresponding Fock-like space can be expressed as a sum of the objects restricted to elementary cells (directed polyhedral graphs) that constitute the lattice. In this case, the classically constructed lattice must have a so-determined structure that includes the projectors into equivalence classes of graphs under all groups of gauge symmetries. These projectors, implemented at the boundaries of elementary cells, will connect the small regions of different quantum geometries.

The second advantage is of an aesthetic nature. The formula in \eqref{holonomy_improvement}, and even more so its holonomies contribution, has an analogous form to the expression that regularizes the Ashtekar connection curvature --- compare the left-hand sides of \eqref{loop_holonomy_symmetrized} and \eqref{holonomy_difference}.


\section{Summary of results}\label{Sec_Summary}

\noindent
The established formulation of canonical LQG is based on a nonunique regularization method. In consequence, it leads to different, hence questionable predictions. Moreover, the approximations assumed in this formulation are not implemented in a consistent, equivalently accurate manner. Finally, the transition from a continuous system to a lattice framework does not agree with the symmetry of the lattice-smeared quantities. This last issue requires the introduction of statistical coarse-graining methods. They are necessary to restore the continuous distribution of observables from large collections of interconnected links- and nodes-distributed lattice-defined operators.

The revision of all the aforementioned shortcomings led to the formulation of this article. The consistent and properly symmetrized procedure of the lattice regularization, which appears to be unique, has been proposed. Moreover, the introduced method has provided a surprisingly symmetric structure of the formulas that involve the smearing along links and around loops. Furthermore, the resulting domain of the regularized Hamiltonian constraint has taken the form of the sum over elementary cells, precisely
\begin{subequations}
\label{regularized_scalar}
\begin{align}
\begin{split}
-\frac{4\sigma}{\gamma\kappa^2}
\sum_v&\,N\pos{v}\.\epsilon^{pqr}
\.\text{tr}
\Big[
\big(h_{qr}\pos{v}-h_{qr}^{-1}\pos{v}\big)
\\
\times&\,
\big\{\mathbf{V}(R\.\pos{v}),\big(h^{\mathstrut}_p\pos{v}-h_p^{-1\mathstrut}\pos{v}\big)\big\}
\Big]\,,
\end{split}
\intertext{or equivalently}
\begin{split}
\frac{8\sigma}{\gamma\kappa^2}
\sum_v&\,N\pos{v}\.\epsilon^{pqr}
\.\text{tr}
\Big[
\big(h_{qr}\pos{v}-h_{qr}^{-1}\pos{v}\big)\tau^i
\Big]
\\
\times&\,
\text{tr}
\Big[
\big\{\mathbf{V}(R\.\pos{v})\big\},\big(h^{\mathstrut}_p\pos{v}-h_p^{-1\mathstrut}\pos{v}\big)\tau^i
\Big]\,.
\end{split}
\end{align}
\end{subequations}
Here, the integral has been formally replaced by the equivalent Riemann sum,
\begin{align}
\label{integral_sum}
\int\!\!d^3x\.f(x)=\lim_{{\bar{l}\to0}}\sum_vf(R\.\pos{v})\.\bar{l}^{\,3}\pos{v}\,,
\end{align}
where the region ${R\.\pos{v}}$ is assumed to be specified by the position of the node ${v\in\partial R\.\pos{v}}$, uniquely located at its boundary. It is worth noting that the corrections of order $\varepsilon$ are not present in the summed elements in \eqref{regularized_scalar} as they occur in the outcome of the standard approach \cite{Thiemann:1996aw,Thiemann:2007zz}. The elements of order $\varepsilon^2$ and higher are neglected, and at the classical level they vanish in the limit ${\bar{l}\to0}$. This limit, determined by a one-dimensional isotropic quantity
${\bar{l}\.\pos{v}}:={\mathbb{L}_0\.\bar{\varepsilon}\.\pos{v}}$,
can be constructed, for instance, by defining ${\bar{\varepsilon}^{\,3}\pos{v}}:=\epsilon^{(\@p\@)\@(\@q\@)\@(\@r\@)}{\varepsilon_{(\@p\@)}\pos{v}}{\varepsilon_{(\@q\@)}\pos{v}}{\varepsilon_{(\@r\@)}\pos{v}}$.

The derived outcome in \eqref{regularized_scalar} is the lattice-smeared analog of the so-called Euclidean term in the Hamiltonian constraint ${\frac{1}{\kappa}\!\int\!d^3xN|E|^{-\frac{1}{2}}\epsilon^{ijk}F^i_{ab}E_j^aE_k^b}$. The second, the so-called Lorentzian term ${-2\frac{\gamma^2\@+1}{\kappa}\!\int\!d^3xN|E|^{-\frac{1}{2}}K^i_aK^j_bE_{[i}^aE_{j]}^b}$, can be calculated from the result in \eqref{regularized_scalar} in the standard exact method \cite{Thiemann:1996ay,Thiemann:1996aw,Thiemann:2007zz}. Let it be also emphasized that the procedure described in this summary can also be implemented regarding any, not only tetrahedral, structure of the lattice. For instance, a quadrilaterally hexahedral lattice provides an analogous outcome, which differs by a constant and the arrangement of elementary cells \cite{Bilski:2021hrr}.

Finally, before the quantization of the lattice Hamiltonian constraint, the densitized dreibein degrees of freedom encoded in ${\mathbf{V}(R\.\pos{v})}$ must be expressed in terms of fluxes through some surfaces. To properly represent the three-dimensional region ${R\.\pos{v}}$ in terms of (a square root of the product of three) two-dimensional fluxes, their probability densities should be integrated within the boundaries of ${R\.\pos{v}}$. The simplest, dimensionally correct orientation of this integral is given along the normal to the two-dimensional fluxes-related surfaces. Considering this three-dimensional continuous distribution of fluxes inside elementary cells of the lattice, the holonomy-flux algebra may lead to different results than the ones only concerning the fluxes at boundaries. Precisely, by replacing $\text{f}[E^a\pos{v}]$ in \eqref{Poisson_holonomy_1} or \eqref{Poisson_holonomy_2} with an integral of the flux continuous distribution between the endpoints of a link, one obtains corrections to an analogous expression with the flux determined only at these endpoints \cite{Bilski:2021hrr,Bilski:2021ysc}. However, by choosing the improved structure of holonomies given in \eqref{holonomy_improvement} and verifying the equivalent algebraic relation, these corrections vanish \cite{Bilski:2021ysc}. This provides yet another argument supporting the improved implementation of the holonomy representation introduced in this article.


\section{Concluding remarks}\label{Sec_Remarks}

\noindent
Let three more facts be emphasized at the end. Despite the criticism of the established regularization procedure, the correction terms of order $\varepsilon$ also vanish in the limit ${\bar{l}\to0}$, although slower than the ones of order $\varepsilon^2$. However, this problem foremostly regards the multiple equivalent choices for the approximation of the Ashtekar connection with a holonomy functional. Moreover, all these choices are one order less accurate than the approximation of the curvature. Consequently, all different Hamiltonian constraint operators would lead to (different) quantum corrections, the structure of which would be more laden with the remnants of a selected regularization. Furthermore, this quantum system would not satisfy Wigner's theorem.

The second issue concerns the transition in \eqref{integral_sum}. This expression is correct when the differential (form) $d^3x$ represents an infinitesimal volume (element) in the Cartesian coordinates. Hence, this relation is exact when ${R\.\pos{v}}$ represents a cuboid and is approximate for an almost cuboidal quadrilateral hexahedron or a pseudoquadrilateral hexahedron (instead of a tetrahedron or a pseudotriangular tetrahedron). This last remark is the most convincing argument to formulate a physically-related lattice model by using the quadrilaterally hexahedral tessellation of the Cauchy hypersurface, \textit{cf.} \cite{Bilski:2021hrr}. Despite this observation, the analysis in this article was carried out by applying the well-established tetrahedral honeycomb (the triangulation or pseudotriangulation) on purpose, but also to demonstrate in which steps the associated problems appear.

To formulate a canonically quantizable model smeared over the triangulated manifold, one should first transform the coordinate system $d^3x$ in \eqref{A_scalar} into the system reflecting the tetrahedral symmetry of ${R\.\pos{v}}$ in \eqref{integral_sum}. This would lead to a different form of the Hamiltonian constraint and dreibeins. Hence, it would require a new analysis of the related symplectic structure, therefore constructing a new regularization procedure. This issue remains open and is worthy of an individual investigation.

The last of the three remarks is ahead of the analyses and results given in this article as it concerns the Fock-like space construction. In the currently established formulation of the gauge-invariant Hilbert spaces over links \cite{Thiemann:2007zz}, the irreducible spin-$j$ representations of holonomies are expressed by the Wigner $D^j(h)$ matrices \cite{Wigner:1931}. The Hamiltonian constraint operator, constructed in terms of the holonomy differences in \eqref{holonomy_difference}, is expected to simplify the research toward the correct vacuum state indication. The natural candidate for this state appears to be $D^j(\mathds{1})=D^j(h_0)$, where $h_0:=h[A=0]$.


\section*{Acknowledgements}

\noindent
This work was partially supported by the National Natural Science Foundation of China grants Nos. 11675145 and 11975203.
The author thanks Piotr Latasiewicz for language editing in the first version of the manuscript.




\end{document}